\documentstyle[amsmath,amssymb,graphicx]{article}

\def\be{\begin{eqnarray}}
\def\ee{\end{eqnarray}}
\def\nn{\nonumber}

\def\p{\partial}

\def\ooplus{\ominus}  

\hoffset= - 1.0in         

\begin{document}

\hfill ITEP/TH-44/12

\bigskip

\centerline{\Large{Introduction to Khovanov Homologies.
}}
\centerline{\Large{II. Reduced
Jones superpolynomials
}}

\bigskip

\centerline{V.Dolotin and A.Morozov}

\bigskip

\centerline{\it ITEP, Moscow, Russia}

\bigskip

\centerline{ABSTRACT}

\bigskip

{\footnotesize

A second part of detailed elementary introduction into Khovanov homologies.
This part is devoted to {\it reduced} Jones superpolynomials.
The story is still about a hypercube of resolutions of a link diagram.
Each resolution is a collection of non-intersecting cycles,
and one associates a 2-dimensional vector space with each cycle.
{\it Reduced} superpolynomial arises when for all cycles, containing
a "marked" edge of the link diagram, the vector space is reduced to
1-dimensional.
The rest remains the same.
Edges of the hypercube are associated with cut-and-join operators,
acting on the cycles.
Superpartners of these operators can be combined into differentials of
a complex, and superpolynomial is the Poincare polynomial of this complex.
HOMFLY polynomials are practically the same in reduced and unreduced case,
but superpolynomials are essentially different,
already in the simplest examples of trefoil and figure-eight knot.
}

\bigskip

\tableofcontents

\newpage

This text is a continuation of \cite{DM}, and is formulated
not independently, but rather as a set of comments to that one.
We assume that the reader is familiar with \cite{DM},
use the same terminology and constructions, without
going into lengthy explanations.
All the references to relevant original works can be also
found in \cite{DM}.
References to formulas from \cite{DM} are given in the form
(I.x) -- in the present paper we refer to the version 1 (v1)
of \cite{DM}.

For relevant foundations of knot theory see \cite{CS}-\cite{Kauf}.
For basic references on Khovanov-Rozansky homologies
and superpolynomials see \cite{Khfirst}-\cite{Khlast}
and \cite{spfirst}-\cite{splast} respectively.

\section{From unreduced to reduced Jones superpolynomial}

\subsection{Ordinary Jones}

The unreduced Jones polynomial in the form (I.13) is obviously
divisible by $D=q+q^{-1}$:
\be
J^{\Gamma_c}(q) = (-)^{n_\circ}q^{n_\bullet-2n_\circ} \!\!\!\!\!\!
\sum_{{\rm resolutions} \ r\ {\rm of}\ \Gamma} \!\!\!\!\!\!
(-q)^{|r-r_c|} D^{\nu_r}
= D\cdot \underline{{ J}}^{\Gamma_c}(q)
\label{bfJ}
\ee
The ratio $\underline{{ J}}^\Gamma(q)$ is called {\it reduced} Jones polynomial.
In what follows we underline the
variables (times and their superpartners), which are being reduced,
as well as the
objects (differential, cohomologies, knot polynomials)
obtained {\it in result} of the reduction
-- hopefully, this does not cause confusion.

To obtain $\underline{{ J}}^\Gamma(q)$ from (I.13) one should say that
one of the cycles at each vertex of the resolutions hypercube
contributes not $D$, but just $1$.
For this purpose one can mark one {\it edge} $E=e$ of $\Gamma$
-- in each resolution there will be exactly one cycle, containing $e$,--
and let {\it this} cycle contribute $1$ instead of $D$.

Since there is always an item in the sum (\ref{bfJ}) with
$\nu_r=1$, there is only one common power of $D$ -- and thus only
one edge can be marked in $\Gamma$ in above construction.
Of course, the answer does not depend on the choice of this single $e$.

\subsection{Reduced superpolynomial}

Now it is clear, what should be done in Khovanov's $T$-deformation:
one should reduce the vector spaces $V$ in (I.33), associated with all the cycles,
which contain the marked edge $e$, from two- to one-dimensional.
This means that the corresponding $\theta$-variables in (I.46) should be nullified.

It is important here that the maps $Q$ in (I.40) diminish the $q$-grading by one:
because of this $\theta$-derivatives are always multiplied by $\theta$ --
what makes such reduction self-consistent. This would not be true if we
tried to nullify $\eta$-variables instead.

In other words, nullification of underlined elements $v_-$ in the following
formulas is self-consistent, while it would not be like that if one
attempts to nullify $v_+$:
\be
\begin{array}{clcc}
\underline{V} \rightarrow \underline{V}\otimes V &&&\\
&v_+ \longrightarrow v_+\otimes v_- + \underline{v_-}\otimes v_+ &
\Longrightarrow & v_+ \longrightarrow v_+\otimes v_-, \nn \\
&\underline{v_-} \longrightarrow \underline{v_-}\otimes v_-
& \Longrightarrow & 0\longrightarrow 0\\
\underline{V}\otimes V \longrightarrow \underline{V} &&&\\
& \underline{v_-}\otimes v_- \longrightarrow 0 & \Longrightarrow
& 0\longrightarrow 0, \\
& \underline{v_-}\otimes v_+ \longrightarrow \underline{v_-}
& \Longrightarrow & 0\longrightarrow 0, \\
& v_+\otimes v_- \longrightarrow \underline{v_-} & \Longrightarrow
& v_+\otimes v_- \longrightarrow 0, \\
& v_+\otimes v_+ \longrightarrow v_+ & \Longrightarrow
& v_+\otimes v_+ \longrightarrow v_+
\end{array}
\ee


\section{Example: Trefoil $3_1$ in a $2$-strand realization}

\subsection{Hypercube, Jones and the cut-and-join operator}

Extended Jones polynomial (I.7.4.1) is
\be
{\cal J}^{\bullet\bullet\bullet} =
\underline{p_3}p_3'+ t(\underline{p_6''} + \underline{p_6'} + \underline{p_6})
+ t^2 (\underline{p_4}p_2 + p_4'\underline{p_2'} + \underline{p_4''}p_2'')
+ t^3p_2\underline{p_2'}p_2''
\label{Jbbb}
\ee
Underlined are the terms, eliminated by reduction
-- and there is exactly one such factor in each item.

This Jones polynomial and the cut an join operator below
are build with the help of the hypercube quiver:

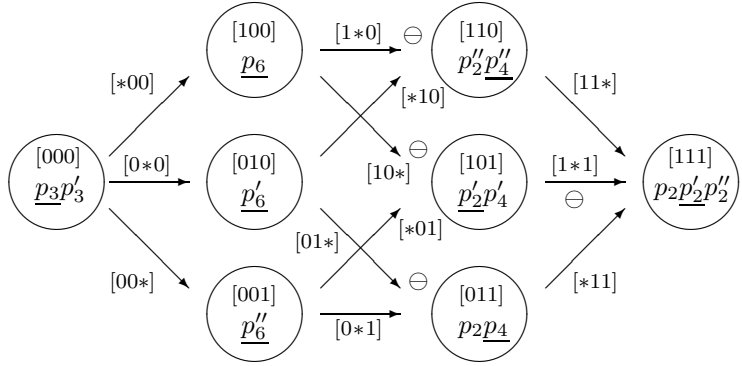
\begin{picture}(200,180)(-100,-80)

\put(-8,7){\mbox{{\footnotesize $[000]$}}}
\put(-8,-5){\mbox{$\underline{p_3}p_3'$}}
\put(20,10){\vector(1,1){30}}
\put(20,0){\vector(1,0){30}}
\put(20,-10){\vector(1,-1){30}}
\put(66,55){\mbox{{\footnotesize$[100]$}}}
\put(70,43){\mbox{$\underline{p_6}$}}
\put(66,5){\mbox{{\footnotesize$[010]$}}}
\put(70,-7){\mbox{$\underline{p_6'}$}}
\put(66,-45){\mbox{{\footnotesize$[001]$}}}
\put(70,-57){\mbox{$\underline{p_6''}$}}
\put(100,50){\vector(1,0){30}}
\put(100,40){\vector(1,-1){30}}
\put(100,10){\vector(1,1){30}}
\put(100,-10){\vector(1,-1){30}}
\put(100,-40){\vector(1,1){30}}
\put(100,-50){\vector(1,0){30}}
\put(152,55){\mbox{{\footnotesize$[110]$}}}
\put(152,43){\mbox{$p_2''\underline{p_4''}$}}
\put(152,5){\mbox{{\footnotesize$[101]$}}}
\put(152,-7){\mbox{$\underline{p_2'}p_4'$}}
\put(152,-45){\mbox{{\footnotesize$[011]$}}}
\put(152,-57){\mbox{$p_2\underline{p_4}$}}
\put(185,40){\vector(1,-1){30}}
\put(185,0){\vector(1,0){30}}
\put(185,-40){\vector(1,1){30}}
\put(231,7){\mbox{{\footnotesize$[111]$}}}
\put(226,-5){\mbox{$p_2\underline{p_2'}p_2''$}}
\put(20,35){\mbox{{\footnotesize$[*00]$}}}
\put(25,4){\mbox{{\footnotesize$[0\!*\!0]$}}}
\put(20,-40){\mbox{{\footnotesize$[00*]$}}}
\put(105,54){\mbox{{\footnotesize$[1\!*\!0]$}}}
\put(117,1){\mbox{{\footnotesize$[10*]$}}}
\put(130,30){\mbox{{\footnotesize$[*10]$}}}
\put(90,-25){\mbox{{\footnotesize$[01*]$}}}
\put(129,-20){\mbox{{\footnotesize$[*01]$}}}
\put(105,-58){\mbox{{\footnotesize$[0\!*\!1]$}}}
\put(195,35){\mbox{{\footnotesize$[11*]$}}}
\put(187,4){\mbox{{\footnotesize$[1\!*\!1]$}}}
\put(195,-40){\mbox{{\footnotesize$[*11]$}}}
\put(131,53){\mbox{{$\ominus$}}}
\put(133,10){\mbox{{$\ominus$}}}
\put(133,-40){\mbox{{$\ominus$}}}
\put(192,-9){\mbox{{$\ominus$}}}
\put(0,0){\circle{35}}
\put(75,50){\circle{35}}
\put(75,0){\circle{35}}
\put(75,-50){\circle{35}}
\put(160,50){\circle{35}}
\put(160,0){\circle{35}}
\put(160,-50){\circle{35}}
\put(240,0){\circle{35}}
\end{picture}

Dimension three of the hypercube is the number of vertices in the
knot diagram, these vertices correspond to the three positions
in the triple sequences $\ [\,\cdot\ \cdot\ \cdot\,]$.

Vertices of the hypercube are in one-to-one correspondence
with resolutions of the knot diagram into sets of non-intersecting
cycles.
If one of the two possible resolutions of the crossing is chosen
in a given vertex of the knot diagram, we put $0$ or $1$ into the
binary label $\ [\,\cdot\ \cdot\ \cdot\,]\ $ of the vertex.
Cycles are labeled by $p_k$, where $k$ is the number of
edges of the knot diagram in the cycle, and different cycles of
the same length are labeled by different number of primes.
The sum of labels $k$ in each hypercube vertex is equal to the number of edges
in the knot diagram (in the present case it is $6$).
Underlined are the cycles, which contain the marked edge of
the knot diagram, there is exactly one underlined $p$ at each
hypercube vertex.

Edges of the hypercube are in one-two-one correspondence with the
flips between two possible resolutions at some vertex of the
knot diagram. Thus they connect two hypecube vertices
differing by a single entry in the binary labels -- it stands
at the position of the vertex where the flip takes place
and we put $*$ at this position to label the edges.
A sign is attributed to the edge, to which a factor $-1$
is contributed by all unities in the binary label, which stand
to the left of the $*$, see (I.45). Hypercube edges to which minus sign
is attributed in this way are labeled by $\ominus$.

Hypercubes are the same for all knot diagrams of the same topology,
and do not depend on the coloring of the vertices of the knot
diagram. This affects only the choice "initial" hypercube vertex,
when $0$ is attributed to all black vertices (where ${\cal R}$ stands
in Turaev-Reshetikhin representation of HOMFLY polynomial)
and $1$ -- to all white vertices (where ${\cal R}^{-1}$ stands).

Extended Jones polynomial is just a sum of the entries over
all hypercube vertices, weighted with $t$ in the power, equal to
the number of arrows, containing the vertex with "initial" one.

Similarly the cut-and-join operator is a sum over all edges,
with $\oplus$ signs marking the terms, where differentials
from associated BRST operator will act with the negative signs,
see below.

\bigskip

Thus the cut-and-join operator, {\it dual} to (\ref{Jbbb}), is
$K^{\bullet\bullet\bullet} = \sum_{i=1}^3 K_i$
with
\be
K_1 = (\underline{p_6}+\underline{p_6'}+\underline{p_6''})
\frac{\p^2}{\p \underline{p_3}\p p_3'}, \nn \\
K_2 = p_2\underline{p_4}\left( \frac{\p}{\p\underline{p_6''}}
\ooplus \frac{\p}{\p\underline{p_6'}}\right) +
\underline{p_2'}p_4\left( \frac{\p}{\p\underline{p_6''}}
\ooplus \frac{\p}{\p\underline{p_6}}\right)
+ p_2''\underline{p_4''}\left( \frac{\p}{\p\underline{p_6'}}
\ooplus \frac{\p}{\p\underline{p_6}}\right), \nn \\
K_3 = \underline{p_2'}p_2''\frac{\p}{\p \underline{p_4}}
\ooplus p_2p_2''\frac{\p}{\p p_4'}
+ p_2\underline{p_2'}\frac{\p}{\p \underline{p_4''}}
\ee
Note that whenever all derivatives over underlined variables
are multiplied by underlined variables -- this will make reduction
procedure consistent for the differentials.
Note also that there are terms in the cut-and-join operator,
where no variables are underlined:
reduction touches all the vertices, but not all the edges of the hypercube.

\subsection{Differentials and cohomologies}

Differentials are naturally written in \cite{DM} through Grassmannian variables
$\vartheta_k = (\theta_k,\eta_k)$, and reduction implies that
the underlined variables $\underline{\theta_k}$ and derivatives over them
are eliminated, leaving only the corresponding $\eta_k$ (which we do not
underline after that).
However, introduction of Grassmannian variables requires accurate work with the
signs. We fix the basis vectors in the space of Grassmannian variables
lexicographically:
$\theta_2'$ stands before $\eta_2'$, before $\theta_2''$, before $\theta_3'$...,
i.e. the basis vector is $\eta_2''\theta_3$ rather than $\theta_3\eta_2''$
or $\theta_3\eta_3'$ rather than $\eta_3'\theta_3$.
Note that the sign $\ooplus$ refers to the sign in the transformation
matrix, the sign of the corresponding term in the differential depends on
the way the variables are ordered in this term -- and does not need to
coincide with $\ooplus$ in bosonic form of the cut-and-join operators.
The second derivative over Grassmann variables is defined in inverse order, as
$\frac{\p^2}{\p\theta\p\eta} = \frac{\p}{\p\eta}\frac{\p}{\p\theta}$,
so that $\frac{\p^2}{\p\theta\p\eta}\theta\eta = +1$.

\bigskip

For example, the differential
\be
d_1 = (\eta_6+\eta_6'+\eta_6'')\frac{\p^2}{\p\eta_3\p\eta_3'} +
( {\theta_6}+ {\theta_6'}+ {\theta_6''})\left(\frac{\p^2}{\p\eta_3\p \theta_3'} +
\frac{\p^2}{\p {\theta_3}\p \eta_3'}\right)
\ee
acts on the basis vectors of the space ${\cal C}_1$  as follows:
\be
d_1 \ \downarrow \ \begin{array}{c|cccc}
{\cal C}_1 & \theta_3\theta_3' & \theta_3\eta_3' & \eta_3\theta_3' & \eta_3\eta_3' \\
&&&&\\
d{\cal C}_1 & 0 & {\theta_6}+ {\theta_6'}+ {\theta_6''} &
{\theta_6}+ {\theta_6'}+ {\theta_6''} &
\eta_6+\eta_6'+\eta_6''
\end{array}
\ee
i.e. provides basis vectors of ${\cal C}_2$ with all signs positive.

If, however, we take $d_3$ the situation will be different.
The naive superpartner to $K_3$ would be
\be
\theta_2 {\theta_2'}\frac{\p}{\p\theta_4''}
\ooplus\theta_2\theta_2''\frac{\p}{\p\theta_4'}
+ {\theta_2'}\theta_2''\frac{\p}{\p\theta_4}
+(\theta_2\eta_2'+\eta_2 {\theta_2'})\frac{\p}{\p\eta_4''}
\ooplus(\theta_2\eta_2''+\eta_2\theta_2'')\frac{\p}{\p\eta_4'}
+( {\theta_2'}\eta_2''+\eta_2'\theta_2'')\frac{\p}{\p\eta_4}
\ee
However such operator converts a basis vector $\theta_2\theta_4$
into $-\theta_2\theta_2'\theta_2''$, which is minus the basis vector.
At the same time another basis vector $\theta_2'\theta_4'$
is converted into $\oplus \theta_2\theta_2'\theta_2''$, i.e. would be
correct if we substitute $-$ instead of $\ooplus$.
This means that the signs should be changed appropriately:
\be
d_3 = -\left(\theta_2 {\theta_2'}\frac{\p}{\p\theta_4''}
\hat{+}\theta_2\theta_2''\frac{\p}{\p\theta_4'}
+ {\theta_2'}\theta_2''\frac{\p}{\p\theta_4}
+(\theta_2\eta_2'+\eta_2 {\theta_2'})\frac{\p}{\p\eta_4''}
\hat{+}(\theta_2\eta_2''+\eta_2\theta_2'')\frac{\p}{\p\eta_4'}
+ ( {\theta_2'}\eta_2''+\eta_2'\theta_2'')\frac{\p}{\p\eta_4}
\right)
\ee
and it turns out that on the place of $\ooplus$ in this case one
should substitute the same signs as everywhere else:
we put hats over the two terms with $\ooplus$
which would naively have different signs.
In eqs.(\ref{imd2}) and (\ref{kerd3}) below we shall see that
the hatted signs are absolutely necessary to guarantee the
nilpotency property $d_3d_2=0$, in particular that
$\ {\rm Im}(d_2)\subset{\rm Ker}(d_3)$.

\bigskip

After this comment we can return to our main line.
The differential $d_1$ is reduced as follows:
\be
d_1 = (\eta_6''+\eta_6'+\eta_6)\frac{\p^2}{\p\eta_3\p\eta_3'} +
(\underline{\theta_6''}+\underline{\theta_6'}+\underline{\theta_6})
\left(\frac{\p^2}{\p\eta_3\p \theta_3'} +
\frac{\p^2}{\p\underline{\theta_3}\p \eta_3'}\right)
\ \Longrightarrow \ \underline{d_1} =
(\eta_6''+\eta_6'+\eta_6)\frac{\p^2}{\p\eta_3\p\eta_3'}
\ee
Underlined here are the variables which should be nullified in the reduction.
Obviously the only derivative w.r.t. an underlined variables ($\underline{\theta_3}$)
comes multiplied by the underlined variables.

The kernel  and cohomology of $d_1$ are changed by the reductions as follows:
\be
{\rm Ker}(d_1) = \Big\{ \underline{\theta_3}\theta_3,\
\eta_3\theta_3'- \underline{\theta_3}\eta_3'\Big\}
& \Longrightarrow & {\rm Ker}(\underline{d_1}) = \Big\{\eta_3\theta_3'\Big\}, \nn \\
{\rm dim}_q(H_0) = {\rm dim}_q{\rm Ker}(d_1) = q^{-2}+1  & \Longrightarrow &
{\rm dim}_q(\underline{H_0}) = {\rm dim}_q{\rm Ker}(\underline{d_1}) = 1
\ee
and the image of $d_1$ is
\be
{\rm Im}(d_1) = \Big\{
\underline{\theta_6''}+\underline{\theta_6'}+\underline{\theta_6},\
\eta_6''+\eta_6'+\eta_6 \Big\} & \Longrightarrow &
{\rm Im}(\underline{d_1}) = \Big\{\eta_6''+\eta_6'+\eta_6\Big\}
\ee

\bigskip

Similarly, for $d_2$:
\be
d_2 = \theta_2\underline{\theta_4}\left(\frac{\p}{\p\underline{\theta_6''}}
- \frac{\p}{\p \underline{\theta_6'}}\right)
+ \underline{\theta_2'}\theta_4'\left(\frac{\p}{\p\underline{\theta_6''}}
- \frac{\p}{\p \underline{\theta_6}}\right)
+ \theta_2''\underline{\theta_4''}\left(\frac{\p}{\p\underline{\theta_6'}}
- \frac{\p}{\p \underline{\theta_6}}\right) + \nn \\
+ (\underline{\eta_2\theta_4} + \theta_2\eta_4)\left(\frac{\p}{\p\eta_6''}
- \frac{\p}{\p\eta_6'}\right)
+ (\eta_2'{\theta_4'} + \underline{\theta_2'}\eta_4')
\left(\frac{\p}{\p\eta_6''} - \frac{\p}{\p\eta_6}\right)
+ (\eta_2''\underline{\theta_4''} + \theta_2''\eta_4'')
\left(\frac{\p}{\p\eta_6'} - \frac{\p}{\p\eta_6}\right)\nn\\
\Longrightarrow \ \ \
\underline{d_2} = \theta_2\eta_4\left(\frac{\p}{\p\eta_6''}
- \frac{\p}{\p\eta_6'}\right)
+ \eta_2'\theta_4'\left(\frac{\p}{\p\eta_6''} - \frac{\p}{\p\eta_6}\right)
+ \theta_2''\eta_4''\left(\frac{\p}{\p\eta_6'} - \frac{\p}{\p\eta_6}\right)
\ee
so that
\be
{\rm Ker}(d_2) = \Big\{\eta_6''+\eta_6'+\eta_6, \
\underline{\theta_6''}+\underline{\theta_6'}+\underline{\theta_6}\Big\}
& \Longrightarrow &
{\rm Ker}(\underline{d_2}) = \Big\{\eta_6''+\eta_6'+\eta_6\Big\}, \nn \\
H_1 = {\rm Ker}(d_2)/{\rm Im}(d_1) = \emptyset & \Longrightarrow &
\underline{H_1} = {\rm Ker}(\underline{d_2})/{\rm Im}(\underline{d_1})
= \emptyset
\ee
and
\be
{\rm Im}(d_2) = \Big\{ \theta_2\underline{\theta_4}
+\underline{\theta_2'}\theta_4', \
\theta_2''\underline{\theta_4''}+\underline{\theta_2'}\theta_4', \
(\eta_2\underline{\theta_4} + \theta_2\eta_4)
+ (\eta_2'{\theta_4'} + \underline{\theta_2'}\eta_4'),\
(\eta_2''\underline{\theta_4''} + \theta_2''\eta_4'')
+ (\eta_2'{\theta_4'} + \underline{\theta_2'}\eta_4')
\Big\} \nn \\ \Longrightarrow \ \ \
{\rm Im}(\underline{d_2}) = \Big\{\theta_2\eta_4 + \eta_2'\theta_4', \
 \theta_2''\eta_4'' + \eta_2'\theta_4'\Big\}
 \label{imd2}
\ee

\bigskip

Finally, for $d_3$:
\be
-d_3 = \theta_2\underline{\theta_2'}\frac{\p}{\p\theta_4''}
\hat{+}\theta_2\theta_2''\frac{\p}{\p\theta_4'}
+\underline{\theta_2'}\theta_2''\frac{\p}{\p\theta_4}
+(\theta_2\eta_2'+\eta_2\underline{\theta_2'})\frac{\p}{\p\eta_4''}
\hat{+}(\theta_2\eta_2''+\eta_2\theta_2'')\frac{\p}{\p\eta_4'}
+(\underline{\theta_2'}\eta_2''+\eta_2'\theta_2'')\frac{\p}{\p\eta_4}
 \nn \\
\Longrightarrow \ \ \
-\underline{d_3} = \hat{+}\theta_2\theta_2''\frac{\p}{\p\theta_4'} +
\theta_2\eta_2'\frac{\p}{\p\eta_4''}
\hat{+}(\theta_2\eta_2''+\eta_2\theta_2'')\frac{\p}{\p\eta_4'} +
\theta_2''\eta_2'\frac{\p}{\p\eta_4}
\ee
so that
$$
{\rm Ker}(d_3) = \Big\{ \theta_2\underline{\theta_4}
\hat{+} \underline{\theta_2'}\theta_4', \
\theta_2''\underline{\theta_4}'' \hat{+} \underline{\theta_2'}\theta_4', \
\theta_2\eta_4\hat{+}\eta_2'\theta_4'-\eta_2''\underline{\theta_4''},\
\theta_2''\eta_4'' \hat{+} \eta_2'\theta_4'-\eta_2\underline{\theta_4},\
\eta_2\underline{\theta_4} \hat{+} \underline{\theta_2'}\eta_4'
+  \eta_2''\underline{\theta_4''}\Big\}
$$
\vspace{-0.5cm}
\be
 & \boxed{\Longrightarrow} &
{\rm Ker}(\underline{d_3}) = \Big\{ \theta_2\eta_4\hat{+}\eta_2'\theta_4',\
\theta_2''\eta_4''\hat{+}\eta_2'\theta_4',\
\eta_2\eta_4 \hat{+} \eta_2'\eta_4'+ \eta_2''\eta_4''\Big\}, \nn \\
H_2 = {\rm Ker}(d_3)/{\rm Im}(d_2)
= \Big\{\eta_2\underline{\theta_4} \hat{+} \underline{\theta_2'}\eta_4'
+  \eta_2''\underline{\theta_4''}\Big\}
\!\!\!\!\!\!\!\!\!\!\!\!\!\!\!\!\!\!\!\!\!
\nn \\
& \boxed{\Longrightarrow} &
\underline{H_2} = {\rm Ker}(\underline{d_3})/{\rm Im}(\underline{d_2})
= \Big\{\eta_2\eta_4 + \eta_2'\eta_4'+ \eta_2''\eta_4''\Big\}, \nn\\
{\rm dim}_q(H_2) = q^0=1, && {\rm dim}_q(\underline{H_2}) = q^2,
\label{kerd3}
\ee
and
\be
{\rm Im}(d_3) = \Big\{\theta_2\underline{\theta_2'}\theta_2'',\
\eta_2\underline{\theta_2'}\theta_2'',\ \theta_2\eta_2'\theta_2'',\
\theta_2\underline{\theta_2'}\eta_2'',\
\theta_2{\eta_2'}\eta_2'',\  \eta_2\eta_2'\theta_2'',\
\eta_2\underline{\theta_2'}\eta_2''\Big\}
 \nn \\ \Longrightarrow \ \ \
{\rm Im}(\underline{d_3}) =
\Big\{\theta_2\eta_2'\theta_2'',\ \eta_2\eta_2'\theta_2'',\
\theta_2\eta_2'\eta_2''\Big\}
\ee
i.e.
\be
\underline{H_3} = {\rm Coim}(\underline{d_3}) = \Big\{\eta_2\eta_2'\eta_2''\Big\},
 \ \ \ \ \
{\rm dim_q}(\underline{H_3}) = q^3
\ee
and this time this coincides with
\be
{H_3} = {\rm Coim}(\underline{d_3}) = \Big\{\eta_2\eta_2'\eta_2''\Big\},
\ \ \ \ \
{\rm dim_q}({H_3}) = q^3
\ee

In fact the boxed arrow above is not naive.
Note that in variance with the cases of $d_1$ and $d_2$
the dimensions of kernels and images
are not just divided by two when the reduction is performed:
it is enough to say that for $d_3$ these dimensions are odd.
Here we encounter a more serious reshuffling.
It is implied by the fact that
\be
d_3 \Big(\eta_2\eta_4+\eta_2'\eta_4'+\eta_2''\eta_4''\Big)
= 2\eta_2\underline{\theta_2'}\eta_2''
\ee
In the image of $d_3$ this fact is reflected just by nullification of
$\eta_2\underline{\theta_2'}\eta_2''\ \in \ {\rm Im}(d_3)$,
which thus drops away from ${\rm Im}(\underline{d_3})$.
However, in the kernels this looks slightly more involved:
this combination of the $q$-grading level $2$
did not belong to ${\rm Ker}(d_3)$,
but it appears ${\rm Ker}(\underline{d_3})$
and starts contributing to $\underline{H_2}$.
Instead, the terms of $q$-grading $0$, which contributed to $H_2$,
are nullified and
disappear from $\underline{H_2}$.
This has a drastic impact on the form of the two superpolynomial,
unreduced and reduced:
the powers of $q$ in front of $T^2$ will be different in these two cases.

\subsection{The Jones superpolynomial}

In result unreduced Jones superpolynomial is
\be
P^{3_1}(q|T) = q^3\sum_{j=0}^3 (qT)^j\,{\rm dim}_q(H_j) = \nn \\ =
q^3\Big((q^{-2}+1)\cdot(qT)^0 + 0 \cdot (qT)^1 + q^0(qT)^2 + q^3(qT)^3\Big)=
q+q^3 + q^5T^2+q^9T^3
\ee
while the reduced one is
\be
\underline{{ P}}^{3_1}(q|T) =
q^{-1}\cdot q^3\sum_{j=0}^3 (qT)^j\,{\rm dim}_q(\underline{H_j})
=\nn\\=
q^2\Big( 1\cdot (qT)^0 + 0\cdot (qT)^1 + q^2(qT)^2 + q^3(qT)^3\Big)
= \boxed{q^2 + q^6T^2 + q^8T^3}
\label{rePtref}
\ee
and
\be
\boxed{
P^{3_1} =  (q+q^{-1})\underline{{ P}}^{3_1} \ -\  q^7(1+T)\,T^2
}
\ee

The extra factor $q^{-1}$ in the definition of $\underline{{ P}}^{3_1}(q|T)$
compensates for non-trivial
$q$-grading power of the reduced 1-dimensional space (we keep $v_+$
with the grading $q$ as its basis element, while it should be rather
shifted to $1$).
Minus sign in front of the last "correction term" shows that
the {\it unreduced} superpolynomial is "more refined":
the $T$-deformation of reduced Jones, after multiplication by $D/q$,
can be further reduced -- to provide a "smaller"  superpolynomial,
which is a deformation of the unreduced Jones.

\section{Reduced Jones superpolynomial for the figure-eight knot}

\subsection{Braid}

The trefoil can be made not only from the $2$-strand braid
with $3$ crossings, but also from the $3$-strand braid with $4$ crossings.
Remarkably, in Khovanov formalism a similar representation for the
non-torus figure-eight knot $4_1$ differs only by the change of coloring.

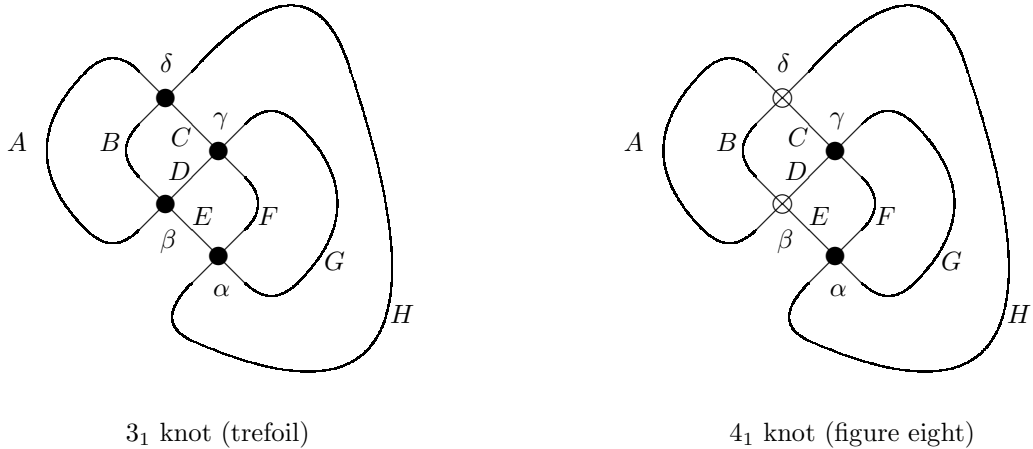
\begin{picture}(200,210)(-100,-110)
\put(0,40){\circle*{7}}
\put(20,20){\circle*{7}}
\put(0,0){\circle*{7}}
\put(20,-20){\circle*{7}}
\put(30,10){\line(-1,1){40}}
\put(30,-30){\line(-1,1){40}}
\put(-10,30){\line(1,1){20}}
\put(30,-10){\line(-1,-1){20}}
\put(-10,-10){\line(1,1){40}}
\qbezier(-10,10)(-20,20)(-10,30)
\qbezier(30,10)(40,0)(30,-10)
%
\qbezier(30,-30)(40,-40)(50,-30)
\qbezier(30,30)(40,40)(50,30)
\qbezier(-10,-10)(-20,-20)(-30,-10)
\qbezier(-10,50)(-20,60)(-30,50)
\qbezier(50,-30)(80,0)(50,30)
\qbezier(-30,-10)(-60,20)(-30,50)
\qbezier(10,50)(55,100)(70,50)
\qbezier(10,-30)(-5,-45)(10,-52)
\qbezier(10,-52)(120,-100)(70,50)
\put(-15,-90){\mbox{$3_1$\ knot (trefoil)}}
\put(18,-35){\mbox{$\alpha$}}
\put(-2,-17){\mbox{$\beta$}}
\put(18,30){\mbox{$\gamma$}}
\put(-2,50){\mbox{$\delta$}}
\put(-60,20){\mbox{$A$}}
\put(-25,20){\mbox{$B$}}
\put(2,22){\mbox{$C$}}
\put(1,9){\mbox{$D$}}
\put(10,-8){\mbox{$E$}}
\put(35,-8){\mbox{$F$}}
\put(60,-25){\mbox{$G$}}
\put(85,-45){\mbox{$H$}}
\end{picture}
\begin{picture}(200,210)(-130,-110)
\put(0,40){\circle{7}}
\put(20,20){\circle*{7}}
\put(0,0){\circle{7}}
\put(20,-20){\circle*{7}}
\put(30,10){\line(-1,1){40}}
\put(30,-30){\line(-1,1){40}}
\put(-10,30){\line(1,1){20}}
\put(30,-10){\line(-1,-1){20}}
\put(-10,-10){\line(1,1){40}}
\qbezier(-10,10)(-20,20)(-10,30)
\qbezier(30,10)(40,0)(30,-10)
\qbezier(30,-30)(40,-40)(50,-30)
\qbezier(30,30)(40,40)(50,30)
\qbezier(-10,-10)(-20,-20)(-30,-10)
\qbezier(-10,50)(-20,60)(-30,50)
\qbezier(50,-30)(80,0)(50,30)
\qbezier(-30,-10)(-60,20)(-30,50)
\qbezier(10,50)(55,100)(70,50)
\qbezier(10,-30)(-5,-45)(10,-52)
\qbezier(10,-52)(120,-100)(70,50)
\put(-20,-90){\mbox{$4_1$\ knot (figure eight)}}
\put(18,-35){\mbox{$\alpha$}}
\put(-2,-17){\mbox{$\beta$}}
\put(18,30){\mbox{$\gamma$}}
\put(-2,50){\mbox{$\delta$}}
\put(-60,20){\mbox{$A$}}
\put(-25,20){\mbox{$B$}}
\put(2,22){\mbox{$C$}}
\put(1,9){\mbox{$D$}}
\put(10,-8){\mbox{$E$}}
\put(35,-8){\mbox{$F$}}
\put(60,-25){\mbox{$G$}}
\put(85,-45){\mbox{$H$}}
\end{picture}

The graph has eight edges and four vertices, labeled by capital latin
and small green letters respectively.
The hypercube is $4$-dimensional and has $16$ vertices:
the graph possesses $16$ resolutions.
These involve $2+4+1+4+4+5 = 20$ different cycles of the lengths $2,3,4,5,6,8$.

Drawing a four-dimensional hypecube is not very informative,
therefore we substitute a picture by a table.
The resolutions (hypecube vertices) are separated
by double horizontal lines into sets with the same $|r-r_c|$ from $r_c=I$,
associated with the $3_1$ knot.
We also explicitly list the vertices, where flips are made to obtain the
resolution from $I$.

\be
\begin{array}{cc|c|c|c|c|c}
{\rm resolution}\ \ &r & {\rm flips\ at} & \multicolumn{3}{c}{{\rm cycles}} & \\
&&& \multicolumn{3}{c}{}& \\
\hline\hline
&\ \ \ \ \ \ \ \ \ &&&&\\
I&  [0000]&& p_2=AB & p_4=CDEH &  \\
  &&& p_2'=FG && \\
&&&&&&\\
\hline\hline
&&&&&&\\
II &[1000] &\alpha& p_2=AB&& p_6=CDEFGH &\\
&&&&&&\\
\hline
&&&&&&\\
III &[0100] &\beta &p_2'=FG&&p_6'=ABDCHE&\\
&&&&&&\\
\hline
&&&&&&\\
IV&[0010] &\gamma& p_2=AB&&p_6''=CGFDEH&\\
&&&&&&\\
\hline
&&&&&&\\
V &[0001] & \delta &p_2'=FG&&p_6'''=ABCDEH &\\
&&&&&&\\
\hline\hline
&&&&&&\\
VI &[1100] &\alpha,\beta&&&&p_8'=ABDCHGFE\\
&&&&&&\\
\hline
&&&&&&\\
VII & [1010] & \alpha,\gamma&p_2=AB& p_3=DEF&&\\
&&&&p_3''=CGH & \\
&&&&&&\\
\hline
&&&&&&\\
VIII &[1001] & \alpha,\delta&&&&p_8=ABCDEFGH\\
&&&&&&\\
\hline
&&&&&&\\
IX & [0110] &\beta,\gamma &&&&p_8''=ABDFGCHE\\
&&&&&&\\
\hline
&&&&&&\\
X & [0101] &\beta,\delta &p_2'=FG&p_3'=AEH&&\\
&&&&p_3'''=BCD&\\
&&&&&&\\
\hline
&&&&&&\\
XI &[0011]&\gamma,\delta&&&&p_8'''=ABCGFDEH\\
&&&&&&\\
\hline\hline
&&&&&&\\
XII &[1110] & \alpha,\beta,\gamma &&p_3''=CGH&p_5''=ABDFE&\\
&&&&&&\\
\hline
&&&&&&\\
XIII &[1101] & \alpha,\beta,\delta&&p_3'''=BCD&p_5'''=AHGFE&\\
&&&&&&\\
\hline
&&&&&&\\
XIV &[1011] &\alpha, \gamma,\delta&&p_3=DEF&p_5=ABCGH&\\
&&&&&&\\
\hline
&&&&&&\\
XV & [0111]&\beta,\gamma,\delta&&p_3'=AEH&p_5'=BCGFD&\\
&&&&&&\\
\hline\hline
&&&&&&\\
XVI &[1111]&\alpha,\beta,\gamma,\delta&&&&p_8''''=AHGCBDFE\\
&&&&&&\\
\end{array}
\nn
\ee

For $4_1$ the starting resolution is different: $r_c= X $,
and the sets are different:
\be
\begin{array}{cc}
|r-r_c| & {\rm resolutions} \ \ r \\
0    & X \\
1 & III,\ V,\  XIII,\ XV \\
2 & I,\ VI,\ VIII,\ IX,\ XI,\ XVI \\
3 & II,\ IV,\  XII,\ XIV \\
4 & VII
\end{array}
\ee

\subsection{Jones polynomials}

From these tables we immediately read the
extended and unreduced Jones polynomials:
\be
{\cal J}^{3_1} =  p_2p_2'p_4 +
t\Big(  p_2p_6+p_2'p_6'+p_2p_6''+p_2'p_6''' \Big) + \nn \\
+ t^2\Big( p_2p_3p_3'' + p_2'p_3'p_3''' + p_8+p_8'+p_8''+p_8''' \Big)
+ t^3\Big(  p_3p_5+p_3'p_5'+p_3''p_5''+p_3'''p_5'''\Big)
+ t^4 p_8''''
\ee
We remind that {\it extended} knot polynomials
are not topological invariants, in particular this ${\cal J}^{3_1}$, defined
through the 3-strand braid, is drastically
different from ${\cal J}^{\bullet\bullet\bullet}$
in (\ref{Jbbb}), defined for the 2-strand braid.
However, the ordinary Jones polynomials, obtained after substitution (I.19)
of $p_k = D$ and $t=-q$ are, of course, the same:
\be
J^{\bullet\bullet\bullet}(q) = J^{3_1}(q)
= q^4\Big(D^3 - 4qD^2 + q^2(2D^3+4D) -4q^3D^2 +q^4D\Big)  =\nn \\ =
D\cdot q^4\Big( D^2-4qD + q^2((2D^2+4) -4q^3D + q^4\Big)
= q+q^3+q^5-q^9 = D\cdot(q^2+q^6-q^8)
\ee

Similarly,
\be
{\cal J}^{4_1} = p_2'p_3'p_3''' + t\Big(p_2'(p_6'+p_6''')
+ p_3'p_5'+p_3'''p_5''' \Big) + \nn \\
+ t^2 \Big( p_2p_2'p_4 +  p_8+p_8'+p_8''+p_8''' + p_8''''\Big)
+ t^3\Big( p_2(p_6+p_6'') + p_3p_5+p_3''p_5'' \Big) + t^4p_2p_3p_3''
\label{4_1extJ}
\ee
and this implies
\be
J^{4_1}(q) = \frac{q^2}{(-q^2)^2}\Big(D^3-4qD^2
+ q^2(D^3+5D) -4q^3D^2 + q^4D^3\Big) = \nn \\ =
D\cdot q^{-2}\Big(D^2 - 4qD+q^2(D^2+5) -4q^3D + q^4D^2\Big)
= q^5+q^{-5} = D\left(q^{-4}-q^{-2}+1-q^{2}+q^{4}\right)
\ee

\subsection{Cut-and-join operators for the $3_1$ knot}

To construct superpolynomials, unreduced and reduced, we need also the differentials.
According to the general procedure, outlined in secs.6 and 7 of \cite{DM},
from (\ref{4_1extJ}) we read the bosonic cut-and-join operators:
\be
\nn\\
K_1 = (p_6+p_6'')\frac{\p^2}{\p p_2'\p p_4} + (p_6'+p_6''')\frac{\p^2}{\p p_2\p p_4}
\nn \\ \nn \\
K_2 = \underbrace{p_3p_3''\left(\frac{\p}{\p p_6''}
\ooplus \frac{\p}{\p p_6}\right)}_{\alpha\gamma}
+ \underbrace{p_3'p_3'''\left(\frac{\p}{\p p_6'''}
\ooplus\frac{\p}{\p p_6'}\right)}_{\beta\delta}
+ \underbrace{ p_8\left(\frac{\p^2}{\p p_2'\p p_6'''}
\ooplus\frac{\p^2}{\p p_2\p p_6}\right)}_{\alpha\delta}
+ \nn \\
+ \underbrace{ p_8'\left(\frac{\p^2}{\p p_2'\p p_6'}
\ooplus\frac{\p^2}{\p p_2\p p_6} \right)}_{\alpha\beta}
+ \underbrace{ p_8''\left(\frac{\p^2}{\p p_2\p p_6''}
\ooplus \frac{\p^2}{\p p_2'\p p_6'} \right)}_{\beta\gamma}
+ \underbrace{ p_8'''\left(\frac{\p^2}{\p p_2'\p p_6'''}
\ooplus\frac{\p^2}{\p p_2\p p_6''} \right)}_{\gamma\delta}
\nn \\ \nn \\ \nn \\
K_3 = \underbrace{p_3''p_5''\left(\frac{\p}{\p p_8'} + \frac{\p}{\p p_8''}\right)
\ooplus p_5''\frac{\p^2}{\p p_2\p p_3} }_{\alpha\beta\gamma}
+ \underbrace{p_3'''p_5'''\left(\frac{\p}{\p p_8'}\ooplus\frac{\p}{\p p_8}\right)
+ p_5'''\frac{\p^2}{\p p_2'\p p_3'} }_{\alpha\beta\delta} + \nn \\
+ \underbrace{p_3p_5\left(\frac{\p}{\p p_8'''}\ooplus\frac{\p}{\p p_8}\right)
+ p_5\frac{\p^2}{\p p_2\p p_3''} }_{\alpha\gamma\delta}
+ \underbrace{p_3'p_5'\left(\frac{\p}{\p p_8''} + \frac{\p}{\p p_8'''}\right)
\ooplus p_5'\frac{\p^2}{\p p_2'\p p_3'''} }_{\beta\gamma\delta}
\nn
\ee
\be
K_4 = p_8''''\left(\ooplus\frac{\p^2}{\p p_3\p p_5} + \frac{\p^2}{\p p_3'\p p_5'}
\ooplus\frac{\p^2}{\p p_3''\p p_5''}+ \frac{\p^2}{\p p_3'''\p p_5'''}\right)
\\ \nn
\label{caj3_1}
\ee
Each item in the cut-and-join operator is associated with an edge of
the hypercube, i.e. with a flip, made at exactly one vertex of original graph.
By $\ooplus$ in this formula we denote the terms where the sign factor
in the super-analogue of the cut-and-join operator will be negative,
$\epsilon^a_{bc}=-1$,  according to the rule (I.45).
There are $12$ (out of $32$) edges of this type.

\subsection{Cut-and-join operators for the $4_1$ knot}

Similarly, for the $4_1$ knot we have:
\be
K_1 = (p_6'''\ooplus p_6')\frac{\p^2}{\p p_3'\p p_3'''}
\ooplus p_5'\frac{\p^2}{\p p_2'\p p_3'''}
+ p_5'''\frac{\p}{\p p_2'\p p_3'}
\nn \\ \nn \\ \nn \\
K_2 = \underbrace{p_2p_4\frac{\p}{\p p_6'} + p_2p_4\frac{\p}{\p p_6'''}}_{I}
+ \underbrace{p_8\left(\frac{\p^2}{\p p_2'\p p_6'''}
\ooplus \frac{\p^2}{\p p_3'''\p p_5'''}\right)}_{\alpha\delta}
+ \underbrace{p_8'\left(\frac{\p^2}{\p p_2'\p p_6'}
+ \frac{\p^2}{\p p_3'''\p p_5'''}\right)}_{\alpha\beta} +
\nn \\
+ \underbrace{p_8''\left(+ \frac{\p^2}{\p p_3'\p p_5'}\right)
\ooplus\frac{\p^2}{\p p_2'\p p_6'}}_{\beta\gamma}
+ \underbrace{p_8'''\left(\frac{\p^2}{\p p_2'\p p_6'''}
+ \frac{\p^2}{\p p_3'\p p_5'}\right)}_{\gamma\delta}
+ \underbrace{p_8''''\left(\frac{\p^2}{\p p_3'\p p_5'}
+ \frac{\p^2}{\p p_3'''\p p_5'''}\right)}_{\alpha\beta\gamma\delta}
\nn \\\nn \\ \nn \\
K_3 = \underbrace{p_6\frac{\p^2}{\p p_2'\p p_4}
\ooplus p_2p_6\left( \frac{\p}{\p p_8} + \frac{\p}{\p p_8'}\right)}_\alpha
\ +\  \underbrace{p_6''\frac{\p^2}{\p p_2'\p p_4}
+ p_2p_6''\left(\frac{\p}{\p p_8''} \ooplus \frac{\p}{\p p_8'''}\right)}_\gamma
+ \nn \\
+ \underbrace{p_3''p_5''\left(\frac{\p}{\p p_8'}
+ \frac{\p}{\p p_8''}  \ooplus \frac{\p}{\p p_8''''}\right)
}_{\alpha\beta\gamma}
+\underbrace{p_3p_5\left(
\frac{\p}{\p p_8'''} \ooplus \frac{\p}{\p p_8} \ooplus \frac{\p}{\p p_8''''}\right)
}_{\alpha\gamma\delta}
\nn \\ \nn \\ \nn \\
K_4 = p_3p_3''\left( \frac{\p}{\p p_6''}\ooplus\frac{\p}{\p p_6}\right)
+p_2p_3''\frac{\p}{\p p_5} \ooplus p_2p_3\frac{\p}{\p p_5''}
\\ \nn
\label{caj4_1}
\ee

Next we convert these bosonic operators into the BRST one,
depending on Grassmannian variables with appropriately chosen signs
in front of the different items to guarantee the nilpotency.
This provides the collection of differentials.

We do not write the lengthy expressions for unreduced differentials,
and proceed directly to  reduced ones, which in the case of $4$-vertex graphs
are still slightly more concise.

\subsection{Reductions}

This time not all the edges are the same, and it looks like at
least two different reductions could be performed, for the marked edges
$A$ (or $B$, $F$, $G$) and $H$ (or $C$, $D$, $E$).

The following table shows by crosses the time-variables which are reduced
(their $\theta$-components eliminated, only $\eta$-components left)
when different edges are marked.
Surviving variables correspond to empty spaces in the table.
The numbers of eliminated and survived variables are shown in the
last two lines respectively.
Clearly, the {\it a priori} (topologically) equivalent reductions
imply nullification of very different $\theta$-variables
(all the $20$ $\eta$'s are always present),
thus the reduced differentials
are different and it is not technically trivial that the resulting cohomologies
are the same.

It is even less trivial that the two {\it a priori} different reductions,
which even leave different numbers of $p$-variables ($8$ and $6$) in the differentials,
give rise to the same reduced superpolynomials.

\be
\begin{array}{c||cc|ccc|cc|c}
& A & B & C & D & E & F & G & H \\
\hline\hline
p_2 = AB & x & x &&&&&&\\
p_2'= FG &&&&&&x&x& \\
\hline
p_3=DEF & &&&x&x&x&&\\
p_3'=AEH & x &&&&x &&& x \\
p_3''=CGH & &&x&&&&x&x\\
p_3'''=BBCD &&x&x&x&&&&\\
\hline
p_4=CDEH & &&x&x&x&&&x\\
\hline
p_5=ABCGH&x&x&x&&&&x&x\\
p_5'=BCGFD& &x&x&x&&x&x&\\
p_5''=ABDFE & x&x&&x&x&x&&\\
p_5'''=AHGFE & x&&&&x&x&x&x\\
\hline
p_6=CDEFGH & &&x&x&x&x&x&x\\
p_6'=ABDCHE&x&x&x&x&x&&&x\\
p_6''=CGFDEH&&&x&x&x&x&x&x\\
p_6'''=ABHEDC&x&x&x&x&x&&&x\\
\hline
{\rm all\ five}\ p_8&x&x&x&x&x&x&x&x\\
\hline\hline
& 12 & 12& 14 & 14& 14& 12&12&14 \\
& 8 & 8 & 6&6&6& 8&8& 6 \\
\end{array}
\nn
\ee

\bigskip

Before proceeding further, we should check the consistency
of our reductions -- that derivatives w.r.t. reduced
variables are always multiplied by reduced variables
and thus do not contribute.
Let us show how this works for the first operator $K_1$ in the
more involved case of $4_1$.
The factor $(p_6'+p_6''')$ is reduced
($\theta_6'$ and $\theta_6'''$ vanish)
in all the reductions,
except for $F$and $G$ -- but in those cases neither
$p_3'$ nor $p_3'''$ is vanishing: this makes all the
reductions of the first item in $K_1$ consistent.
Likewise, in the second item $p_5'$ is reduced
in all cases except for $A$, $E$ and $H$,
but neither $p_2'$ nor $p_3'''$ are reduced.
The same happens in the third item: when $p_5'''$
remains unreduced (in $B$, $C$ and $D$ reductions),
neither $p_2'$ nor $p_3'$ are reduced.
In a similar way for both $4_1$ and $3_1$ one can check self-consistency of
all eight reductions of all the four operators $K$
-- and of their super-counterparts -- the differentials $d$.
All this will be transparently seen in explicit calculations below --
but those we do only for one of the eight: for $A$-reduction.

\section{$A$-reduction of the cut-and-join operator for the $3_1$ knot}

First of all, we write the cut-and-join operators (\ref{caj3_1}),
underlying the reduced variables $p$: this means that the corresponding
Grassmannian variable $\vartheta$ contains only $\eta$-, but not the $\theta$-component.

\subsection{Reduction of bosonic operator (\ref{caj3_1})}
\vspace{-0.3cm}
\be
\underline{K_1} = (p_6+p_6'')\frac{\p^2}{\p p_2'\p p_4} +
(\underline{p_6'}+\underline{p_6'''})\frac{\p^2}{\p \underline{p_2}\p p_4}
\nn \\ \nn \\ \nn \\
\underline{K_2} =  p_3p_3''\left(\frac{\p}{\p p_6''}\ooplus\frac{\p}{\p p_6}\right)
+  \underline{p_3'}p_3'''
\left(\frac{\p}{\p \underline{p_6'''}}\ooplus\frac{\p}{\p \underline{p_6'}}\right)
+   \underline{p_8}
\left(\frac{\p^2}{\p p_2'\p \underline{p_6'''}}
\ooplus\frac{\p^2}{\p \underline{p_2}\p p_6}
\right)
+ \nn \\
+ \underline{p_8'}  \left(\frac{\p^2}{\p p_2'\p \underline{p_6'}}
\ooplus\frac{\p^2}{\p \underline{p_2}\p p_6 }\right)
+   \underline{p_8''}\left(\frac{\p^2}{\p \underline{p_2}\p p_6''}
\ooplus\frac{\p^2}{\p p_2'\p \underline{p_6'}}\right)
+   \underline{p_8'''}\left(\frac{\p^2}{\p p_2' \p \underline{p_6'''}}
\ooplus  \frac{\p^2}{\p \underline{p_2}\p p_6''}  \right)
\nn \\ \nn \\ \nn \\
\underline{K_3} =  p_3''\underline{p_5''} \left(\frac{\p}{\p \underline{p_8'}}
+ \frac{\p}{\p \underline{p_8''}}\right)
\ooplus \underline{p_5''}\frac{\p^2}{\p \underline{p_2}\p p_3}
+  p_3'''\underline{p_5'''}\left(\frac{\p}{\p \underline{p_8'}}
\ooplus\frac{\p}{\p \underline{p_8}}\right)
+ \underline{p_5'''}\frac{\p^2}{\p p_2'\p \underline{p_3'}}  + \nn \\
+  p_3\underline{p_5}\left( \frac{\p}{\p \underline{p_8'''}}
\ooplus\frac{\p}{\p \underline{p_8}}\right)
+ \underline{p_5}\frac{\p^2}{\p \underline{p_2}\p p_3''}
+  \underline{p_3'}p_5'\left(\frac{\p}{\p \underline{p_8''}}
+ \frac{\p}{\p \underline{p_8'''}}\right)
\ooplus p_5'\frac{\p^2}{\p p_2'\p p_3'''}
\nn
\ee
\be
\underline{K_4} = \underline{p_8''''}\left(\ooplus\frac{\p^2}{\p p_3\p \underline{p_5}}
+ \frac{\p^2}{\p \underline{p_3'}\p p_5'} \ooplus
\frac{\p^2}{\p p_3''\p \underline{p_5''}}
+ \frac{\p^2}{\p p_3'''\p \underline{p_5'''}}\right)
\\ \nn \\ \nn
\ee
These formulas allow us to write the $A$-reduced differentials in the $3_1$
case and calculate their cohomologies.

\subsection{Differential \underline{$d_1$}\ and the cohomology \underline{$H_0$}}

From
\be
\underline{K_1} = (p_6+p_6'')\frac{\p^2}{\p p_2'\p p_4} +
(\underline{p_6'}+\underline{p_6'''})\frac{\p^2}{\p \underline{p_2}\p p_4}
\ee
we obtain:
\be
\underline{d_1} = -(\theta_6 + \theta_6'')\left(\frac{\p^2}{\p \theta_2'\p\eta_4} +
\frac{\p^2}{\p \eta_2'\p\theta_4}\right)
-(\eta_6+\eta_6'') \frac{\p^2}{\p \eta_2'\p \eta_4}
+(\eta_6'+\eta_6''')\frac{\p^2}{\p \eta_2\p \eta_4}
\ee
Note non-trivial signs in this formula:
they are such that the $\underline{d_1}$-images of all basis
vectors of the 4-dimensional space
$\underline{{\cal C}_1} = \underline{V_2}\otimes V_2'\otimes  V_4$
(its quantum dimension is ${\rm dim}_q(\underline{{\cal C}_1}) = qD^2$)
are expressed through the basis vectors of
$\underline{{\cal C}_2}$ with non-negative coefficients
(we remind our convention $\frac{\p^2}{\p\theta\p\eta} (\theta\eta) = +1$):

\bigskip

\centerline{
$
\underline{d_1}\ \downarrow\
\begin{array}{c||c|c|c|c|}
\underline{{\cal C}_0} &
\eta_2\theta_2'\theta_4&  \eta_2\theta_2'\eta_4&
\eta_2\eta_2'\theta_4& \eta_2\eta_2'\eta_4\\
&&&&\\
\underline{d_1}\underline{{\cal C}_0} &
0 & \eta_2(\theta_6+\theta_6'')+\theta_2'(\eta_6'+\eta_6''')
& \eta_2(\theta_6+\theta_6'')
&\eta_2(\eta_6+\eta_6'') + \eta_2'(\eta_6'+\eta_6''')
\end{array}
$
}

\bigskip

\noindent
Clearly,
\be
{\rm Ker}(\underline{d_1}) = \{\eta_2\theta_2'\theta_4\},
\ \ \ \ {\rm dim}_q(\underline{H_0}) =
{\rm dim}_q{\rm Ker}(\underline{d_1}) = q^{-1}
\ee
The image
\be
{\rm Im}(\underline{d_1}) =
\Big\{ \eta_2(\theta_6 + \theta_6''),\ \theta_2'(\eta_6'+\eta_6'''),\
\eta_2(\eta_6+\eta_6'') + \eta_2'(\eta_6'+\eta_6''')\Big\}, \ \ \ \ \
{\rm dim}_q{\rm Im}(\underline{d_1}) = 2+q^2
\ee
Note that differential itself is not really needed in this calculation:
what we need are the vector spaces with basises and bosonic cut-and-join operator
with the $\ooplus$ labels -- this is enough to construct the kernel
and image of the differential, without knowing its explicit form
(the sign assignments).

\subsection{Differential \underline{$d_2$}\ and the cohomology \underline{$H_1$}}

This time we begin with
\be
\underline{K_2} =  p_3p_3''\left(\frac{\p}{\p p_6''}\ooplus\frac{\p}{\p p_6}\right)
+  \underline{p_3'}p_3'''
\left(\frac{\p}{\p \underline{p_6'''}}\ooplus\frac{\p}{\p \underline{p_6'}}\right)
+   \underline{p_8}
\left(\frac{\p^2}{\p p_2'\p \underline{p_6'''}}
\ooplus\frac{\p^2}{\p \underline{p_2}\p p_6}
\right)
+ \nn \\
+ \underline{p_8'}  \left(\frac{\p^2}{\p p_2'\p \underline{p_6'}}
\ooplus\frac{\p^2}{\p \underline{p_2}\p p_6 }\right)
+   \underline{p_8''}\left(\frac{\p^2}{\p \underline{p_2}\p p_6''}
\ooplus\frac{\p^2}{\p p_2'\p \underline{p_6'}}\right)
+   \underline{p_8'''}\left(\frac{\p^2}{\p p_2' \p \underline{p_6'''}}
\ooplus  \frac{\p^2}{\p \underline{p_2}\p p_6''}  \right)
\ee
The signs in the differential $\underline{d_2}$ are now adjusted in two steps:
first, we choose the signs so that basis vectors are mapped into basis vectors
with non-negative coefficients, second we reverse the signs in the terms,
marked by the $\ominus$ signs. As usual these terms will be marked by a hat.
In this way we get:
\be
\underline{d_2} =
-\theta_3\theta_3''\left(\frac{\p}{\p\theta_6''}\,\hat{-}\frac{\p}{\p\theta_6}\right)
-(\theta_3\eta_3''+\eta_3\theta_3'')
\left(\frac{\p}{\p \eta_6''}\,\hat{-}\frac{\p}{\p\eta_6}\right)
-   {\eta_3'}\theta_3'''
\left(\frac{\p}{\p  {\eta_6'''}}\, \hat{-} \frac{\p}{\p  {\eta_6'}}\right)
+ \nn \\
+    {\eta_8}\left(\frac{\p^2}{\p \eta_2'\p \eta_6'''}
\, \hat{-} \frac{\p^2}{\p \eta_2\p  {\eta_6}}\right)
+  {\eta_8'}  \left( \frac{\p^2}{\p\eta_2'\p \eta_6' }
\,\hat{-} \frac{\p^2}{\p \eta_2\p  {\eta_6}}\right) + \nn \\
+   {\eta_8''}\left(\frac{\p^2}{\p  \eta_2\p \eta_6''}
\,\hat{-} \frac{\p^2}{\p \eta_2'\p  \eta_6'}
\right)
+ {\eta_8'''}\left(  \frac{\p^2}{\p \eta_2'  \p  \eta_6'''}
\,\hat{-}\frac{\p^2}{\p  {\eta_2}\p \eta_6''}\right)
\ee
which acts on the space ${\cal C}_2$ as follows:

\bigskip

\centerline{
$
\underline{d_2}\ \downarrow\
\begin{array}{c||c|c|c|c|c|c|c|c|}
\underline{{\cal C}_1} & \eta_2\theta_6 & \eta_2\eta_6
& \theta_2'\eta_6' &\eta_2'\eta_6'
& \eta_2\theta_6'' & \eta_2\eta_6''
& \theta_2'\eta_6''' & \eta_2'\eta_6'''\\
&&&&&&&&\\
\underline{d_2}\underline{{\cal C}_1} &
\hat{-}\eta_2\theta_3\theta_3''
& \hat{-}\eta_2(\theta_3\eta_3''+\eta_3\theta_3'') &
\hat{-}\theta_2'\eta_3'\theta_3''' & \hat{-}\eta_2'\eta_3\theta_3'''
& \eta_2\theta_3\theta_3'' & \eta_2(\theta_3\eta_3''+\eta_3\theta_3'')
& \theta_2'\eta_3'\theta_3''' & \eta_2'\eta_3'\theta_3'''\\
&& \hat{-}\eta_8\hat{-}\eta_8' && +\eta_8'\hat{-}\eta_8'' &&
+\eta_8''\hat{-}\eta_8''' &&+\eta_8 + \eta_8'''
\end{array}
$
}

\bigskip

\noindent
Note that all the minus signs in the target space appear with hats:
this is the rule to ascribe signs to the items in the differential.

\bigskip

\noindent
It follows that
\be
{\rm Ker}(\underline{d_2}) = \Big\{ \eta_2(\theta_6+\theta_6''),\ \
\theta_2'(\eta_6'+\eta_6'''),\ \
\eta_2(\eta_6+\eta_6'') + \eta_2'(\eta_6'+\eta_6''')\Big\},
\ \ \ \ \ \ \ {\rm dim}_q{\rm Ker}(\underline{d_2}) = 2+q^2\nn \\
\underline{H_1} = {\rm Ker}(\underline{d_2})/{\rm Im}(\underline{d_1})
= \emptyset
\ee
and
\be
{\rm Im}(\underline{d_2}) = \Big\{ \eta_2\theta_3\theta_3'',\ \
\theta_2'\eta_3'\theta_3'',\ \
\eta_2(\theta_3\eta_3''+\eta_3\theta_3'') + \eta_8+ \eta_8',\ \
\eta_2'\eta_3'\theta_3''' + \eta_8 + \eta_8''', \ \
\eta_8+\eta_8'-\eta_8''+\eta_8'''
 \Big\},\nn \\
{\rm dim}_q{\rm Im}(\underline{d_2}) =  2q^{-1} + 3q
\ee

\subsection{Differential \underline{$d_3$}\ and the cohomology \underline{$H_2$}}

Starting from
\be
\underline{K_3} =  p_3''\underline{p_5''} \left(\frac{\p}{\p \underline{p_8'}}
+ \frac{\p}{\p \underline{p_8''}}\right)
\ooplus \underline{p_5''}\frac{\p^2}{\p \underline{p_2}\p p_3}
+  p_3'''\underline{p_5'''}\left(\frac{\p}{\p \underline{p_8'}}
\ooplus\frac{\p}{\p \underline{p_8}}\right)
+ \underline{p_5'''}\frac{\p^2}{\p p_2'\p \underline{p_3'}}  + \nn \\
+  p_3\underline{p_5}\left( \frac{\p}{\p \underline{p_8'''}}
\ooplus\frac{\p}{\p \underline{p_8}}\right)
+ \underline{p_5}\frac{\p^2}{\p \underline{p_2}\p p_3''}
+  \underline{p_3'}p_5'\left(\frac{\p}{\p \underline{p_8''}}
+ \frac{\p}{\p \underline{p_8'''}}\right)
\ooplus p_5'\frac{\p^2}{\p p_2'\p p_3'''}
\ee
we obtain by the same procedure:
\be
\underline{d_3} =
 \theta_3''{\eta_5''} \left(\frac{\p}{\p {\eta_8'}}
+ \frac{\p}{\p  {\eta_8''}}\right)
\hat{+}\,  {\eta_5''}\frac{\p^2}{\p  {\eta_2}\p \eta_3}
+  \theta_3''' {\eta_5'''}\left(\frac{\p}{\p  {\eta_8'}}
\hat{-}\frac{\p}{\p  {\eta_8}}\right)
- {\eta_5'''}\frac{\p^2}{\p \eta_2'\p  {\eta_3'}}  + \nn \\
+\theta_3 {\eta_5}\left(\frac{\p}{\p  {\eta_8'''}}
\hat{-}\frac{\p}{\p {\eta_8}}\right)
+  {\eta_5}\frac{\p^2}{\p  {\eta_2}\p \eta_3''}
+   {\eta_3'}\theta_5'\left(\frac{\p}{\p  {\eta_8''}}
+ \frac{\p}{\p  {\eta_8'''}}\right)
\hat{-}\,\theta_5'\left(\frac{\p^2}{\p \theta_2'\p\eta_3'''}
+ \frac{\p^2}{\p\eta_2'\p\theta_3'''}\right)
\hat{-}\, \eta_5'\frac{\p^2}{\p \eta_2'\p \eta_3'''}
\ee

\bigskip

\centerline{
$
\underline{d_3}\ \downarrow\
\begin{array}{c||c|c|c|c|c|c|c|c|}
\underline{{\cal C}_2} & \eta_2\theta_3\theta_3'' & \eta_2\theta_3\eta_3'' &
\eta_2\eta_3\theta_3'' & \eta_2\eta_3\eta_3'' & \theta_2'\eta_3'\theta_3'''
& \theta_2'\eta_3'\eta_3''' & \eta_2'\eta_3'\theta_3''' & \eta_2'\eta_3'\eta_3'''\\
&&&&&&&&\\
\underline{d_3}\underline{{\cal C}_2} &
0 & \theta_3\eta_5 & \hat{-}\theta_3''\eta_5'' & \eta_3\eta_5\hat{-}\eta_3''\eta_5''
& 0 & \hat{-}\eta_3'\theta_5' & \hat{-}\eta_3'\theta_5'+\theta_3'''\eta_5'''
& \hat{-}\eta_3'\eta_5'+\eta_3'''\eta_5'''
\end{array}
$
}

\bigskip

\centerline{
$
\begin{array}{|c|c|c|c|}
\eta_8 & \eta_8' & \eta_8'' & \eta_8''' \\
&&&\\
\hat{-}\theta_3\eta_5\,\hat{-}\theta_3'''\eta_5'''
& \theta_3''\eta_5'' + \theta_3'''\eta_5'''
& \eta_3'\theta_5' + \theta_3''\eta_5'' & \theta_3\eta_5 + \eta_3'\theta_5'
\end{array}
$
}

\bigskip

\be
{\rm Ker}(\underline{d_3}) = \Big\{\eta_2\theta_3\theta_3'',\ \
\theta_2'\eta_3'\theta_3''',\ \
\eta_8+\eta_8' + \eta_2(\theta_3\eta_3''+\eta_3\theta_3''),\nn \\
\eta_8+\eta_8'''+\eta_2'\eta_3'\theta_3''',\ \
\eta_2\eta_3\theta_3''+\theta_2'\eta_3'\eta_3'''+\eta_8'',\ \
\eta_8+\eta_8'-\eta_8''+\eta_8'''
\Big\},\nn \\
{\rm dim}_q{\rm Ker}(\underline{d_3}) = 2q^{-1}+4q, \nn \\
\underline{H_2} = {\rm Ker}(\underline{d_3})/{\rm Im}(\underline{d_2})
=  \Big\{ \eta_2\eta_3\theta_3''+\theta_2'\eta_3'\eta_3'''+\eta_8'' \Big\},
\ \ \ \ \ \ \ {\rm dim}_q(\underline{H_2}) = q
\ee
and
\be
{\rm Im}(\underline{d_3}) = \Big\{\theta_3\eta_5,\ \ \theta_5'\eta_3',\ \
\theta_3''\eta_5'',\ \ \theta_3'''\eta_5''',\ \ \eta_3\eta_5-\eta_3''\eta_5'',\ \
\eta_3'\eta_5'-\eta_3'''\eta_5''' \Big\}
\ee

\subsection{Differential \underline{$d_4$}\ and the cohomologies \underline{$H_3$}\
and \underline{$H_4$}}

Finally, from
\be
\underline{K_4} = \underline{p_8''''}
\left(\ooplus\frac{\p^2}{\p p_3\p \underline{p_5}}
+ \frac{\p^2}{\p \underline{p_3'}\p p_5'} \ooplus
\frac{\p^2}{\p p_3''\p \underline{p_5''}} +
\frac{\p^2}{\p p_3'''\p \underline{p_5'''}}\right)
\ee

\be
\underline{d_4} = \eta_8''''\left(\hat{-}\frac{\p^2}{\p \eta_3\p\eta_5} +
\frac{\p^2}{\p \eta_3'\p\eta_5'}\, \hat{-} \frac{\p^2}{\p \eta_3''\p\eta_5''} +
\frac{\p^2}{\p \eta_3'''\p\eta_5'''}\right)
\\ \nn
\ee
Thus in the $4\cdot 2 =8$-dimensional space
$C_4 = V_3\otimes\underline{V_5}\, \oplus\, \underline{V_3}\otimes{V_5}\,
\oplus\, V_3\otimes\underline{V_5}\, \oplus\, V_3\otimes\underline{V_5}$
the $7$-dimensional kernel
\be
{\rm ker}(\underline{d_4}) = \Big\{\theta_3\eta_5,\ \eta_3'\theta_5',\
\theta_3''\eta_5'',\ \theta_3'''\eta_5''',\ \eta_3\eta_5+\eta_3'\eta_5', \
\eta_3\eta_5-\eta_3''\eta_5'', \ \eta_3'\eta_5'-\eta_3'''\eta_5'''\Big\}
\ee
Therefore
\be
\underline{H_3} = {\rm Ker}(\underline{d_4})/{\rm Im}(\underline{d_3})
= \Big\{ \eta_3\eta_5+\eta_3'\eta_5' \Big\}
\ \ \ {\rm and}\ \ \  {\rm dim}_q(\underline{H_3}) = q^2
\ee

The space $C_5 = \underline{V}$ is one-dimensional, the basis element is
$\eta_8''''$ and it coincides with the $8-7=1$-dimensional image of $\underline{d_4}$.
Therefore the coimage of $\underline{d_4}$ is empty,
and the cohomology
\be
\underline{H_4}=0
\ee

\subsection{Reduced Jones superpolynomials for $3_1$}

Thus
\be
{\rm dim}_q(\underline{H_0})=q^{-1}, \nn\\
{\rm dim}_q(\underline{H_1})=0, \nn\\
{\rm dim}_q(\underline{H_2})=q, \nn\\
{\rm dim}_q(\underline{H_3})=q^{2}, \nn\\
{\rm dim}_q(\underline{H_4})=0
\ee
and therefore
\be
\underline{P}^{3_1}_\Box =
q^{-1}\cdot q^4 \sum_{i=0}^4 (qT)^i\cdot {\rm dim_q}(\underline{H_i})
= \frac{q^4}{q} \Big(q^{-1} + 0\cdot(qT) + q^1\cdot (qT)^2 + q^2\cdot (qT)^3 \Big)
= \boxed{ q^2 + q^6T^2 + q^8T^3 }
\ee
what coincides with the answer (\ref{rePtref}),
obtained from the $2$-strand representation.

\bigskip

\section{$A$-reduced differentials for the $4_1$ knot}

Now we repeat the same procedure in the case of the figure-eight knot.

\subsection{Reduction of bosonic operator (\ref{caj4_1})}

\be
\underline{K_1} = (\underline{p_6'''} \ooplus\underline{p_6'})
\frac{\p^2}{\p \underline{p_3'}\p p_3'''}
\ooplus p_5'\frac{\p^2}{\p p_2'\p p_3'''}
+ \underline{p_5'''}\frac{\p}{\p p_2'\p \underline{p_3'}}
\nn \\ \nn \\ \nn \\
\underline{K_2} =  \underline{p_2}p_4\frac{\p}{\p \underline{p_6'}}
+ \underline{p_2}p_4\frac{\p}{\p \underline{p_6'''}}
+ \underline{p_8}\left(\frac{\p^2}{\p p_2'\p \underline{p_6'''}}
\ooplus \frac{\p^2}{\p p_3'''\p \underline{p_5'''}}\right)
+ \underline{p_8'}\left(\frac{\p^2}{\p p_2'\p \underline{p_6'}}
+ \frac{\p^2}{\p p_3'''\p \underline{p_5'''}}\right)  +
\nn \\
+ \underline{p_8''}\left(\frac{\p^2}{\p \underline{p_3'}\p p_5'}
\ooplus\frac{\p^2}{\p p_2'\p \underline{p_6'}}\right)
+ \underline{p_8'''}\left(\frac{\p^2}{\p p_2'\p \underline{p_6'''}}
+ \frac{\p^2}{\p \underline{p_3'}\p p_5'}\right)
+ \underline{p_8''''}\left(\frac{\p^2}{\p \underline{p_3'}\p p_5'}
+ \frac{\p^2}{\p p_3'''\p \underline{p_5'''}}\right)
\nn
\ee
\be
\underline{K_3} =  p_6\frac{\p^2}{\p p_2'\p p_4}
\ooplus \underline{p_2}p_6\left(\frac{\p}{\p \underline{p_8}}
+ \frac{\p}{\p \underline{p_8'}}\right)
\ +\   p_6''\frac{\p^2}{\p p_2'\p p_4}
+ \underline{p_2}p_6''\left(\frac{\p}{\p \underline{p_8''}}
\ooplus \frac{\p}{\p \underline{p_8'''}}\right)
+ \nn \\
+  p_3''\underline{p_5''}\left(\frac{\p}{\p \underline{p_8'}}
+ \frac{\p}{\p \underline{p_8''}}
\ooplus \frac{\p}{\p \underline{p_8''''}}\right)
+  p_3\underline{p_5}\left(
\frac{\p}{\p \underline{p_8'''}} \ooplus \frac{\p}{\p \underline{p_8}}
\ooplus \frac{\p}{\p \underline{p_8''''}}\right)
\nn \\ \nn \\ \nn \\
\underline{K_4} = p_3p_3''\left( \frac{\p}{\p p_6''}
\ooplus\frac{\p}{\p p_6}\right)
+\underline{p_2}p_3''\frac{\p}{\p \underline{p_5}}
\ooplus \underline{p_2}p_3\frac{\p}{\p \underline{p_5''}}
\\ \nn
\label{caj4_1red}
\ee

\subsection{Differential \underline{$d_1$}\ and the cohomology \underline{$H_0$}}

As usual, we begin from the relevant bosonic operator
\be
\underline{K_1} = (\underline{p_6'''} \ooplus\underline{p_6'})
\frac{\p^2}{\p \underline{p_3'}\p p_3'''}
\ooplus p_5'\frac{\p^2}{\p p_2'\p p_3'''}
+ \underline{p_5'''}\frac{\p}{\p p_2'\p \underline{p_3'}}
\ee
In this case the space
$\underline{{\cal C}_0} =V_2'\otimes \underline{V_3'}\otimes V_3'''$
has dimension ${\rm dim}_q(\underline{{\cal C}_0}) = q^{-1}+2q+q^3=qD^2$.
The differential
\be
\underline{d_1} =
-(\eta_6'''-\eta_6')\frac{\p^2}{\p\eta_3'\p\eta_3'''}
\hat{-} \theta_5'\left(\frac{\p^2}{\p \theta_2'\p\eta_3'''}
+ \frac{\p^2}{\p\eta_2'\p\theta_3'''}\right)
\hat{-} \eta_5'\frac{\p^2}{\p\eta_2'\p\eta_3'''}
-\eta_5'''\frac{\p^2}{\p\eta_2'\p\eta_3'}
\ee
converts the basis in this space as follows:
\be
\underline{d_1}\ \downarrow\
\begin{array}{c||c|c|c|c|}
\underline{{\cal C}_0} & \theta_2'\eta_3'\theta_3'''&
\theta_2'\eta_3'\eta_3'''&  \eta_2'\eta_3'\theta_3'''&
\eta_2'\eta_3'\eta_3''' \\ &&&&\\
\underline{d_1}\underline{{\cal C}_0} & 0&
\theta_2'(\eta_6'''\hat{-}\eta_6') \hat{-}\eta_3'\theta_5'&
\theta_3'''\eta_5'''\hat{-}\theta_3'\eta_5'&
\eta_2'(\eta_6'''\hat{-}\eta_6') \hat{-} \eta_3'\eta_5'+\eta_3'''\eta_5'''
\end{array}
\ee
Clearly,
\be
{\rm Ker}(\underline{d_1}) = \{\theta_2'\eta_3'\theta_3'''\},
\ \ \ \ \ \ \ {\rm dim}_q(\underline{H_0}) =
{\rm din}_q{\rm Ker}(\underline{d_1}) = q^{-1}
\ee
Note that the second item in $\underline{d_1}$
would annihilate also the linear combination
$\theta_2'\eta_3'\eta_3''' -
\eta_2'\eta_3'\theta_3'''$, but it is not annihilated by
the first item in $\underline{d_1}$. The image
\be
{\rm Im}(\underline{d_1}) = \Big\{
\theta_2'(\eta_6'\hat{-}\eta_6''')+\eta_3'\theta_5',\ \
\theta_3'\eta_5'\hat{-}\theta_3'''\eta_5''',\ \
\eta_2'(\eta_6'\hat{-}\eta_6''')
+ \eta_3'\eta_5' \hat{-} \eta_3'''\eta_5''' \Big\}, \ \ \ \ \ \ \
{\rm dim}_q{\rm Im}(\underline{d_1}) = 2+q^2
\ee

\subsection{Differential \underline{$d_2$}\ and the cohomology \underline{$H_1$}}

From
\be
\underline{K_2} =  \underline{p_2}p_4\frac{\p}{\p \underline{p_6'}}
+ \underline{p_2}p_4\frac{\p}{\p \underline{p_6'''}}
+ \underline{p_8}\left(\frac{\p^2}{\p p_2'\p \underline{p_6'''}}
\ooplus \frac{\p^2}{\p p_3'''\p \underline{p_5'''}}\right)
+ \underline{p_8'}\left(\frac{\p^2}{\p p_2'\p \underline{p_6'}}
+ \frac{\p^2}{\p p_3'''\p \underline{p_5'''}}\right)  +
\nn \\
+ \underline{p_8''}\left(\frac{\p^2}{\p \underline{p_3'}\p p_5'}
\ooplus\frac{\p^2}{\p p_2'\p \underline{p_6'}}\right)
+ \underline{p_8'''}\left(\frac{\p^2}{\p p_2'\p \underline{p_6'''}}
+ \frac{\p^2}{\p \underline{p_3'}\p p_5'}\right)
+ \underline{p_8''''}\left(\frac{\p^2}{\p \underline{p_3'}\p p_5'}
+ \frac{\p^2}{\p p_3'''\p \underline{p_5'''}}\right)
\ee
we get the differential:
\be
\underline{d_2} =
\eta_2\theta_4\left(\frac{\p}{\p \eta_6'} + \frac{\p}{\p\eta_6'''}\right)
+ \eta_8\left(\frac{\p^2}{\p\eta_2'\p\eta_6'''}
\,\hat{-}\frac{\p^2}{\p\eta_3'''\p\eta_5'''}\right)
+ \eta_8'\left(\frac{\p^2}{\p\eta_2'\p\eta_6'}
+\frac{\p^2}{\p\eta_3'''\p\eta_5'''}\right) +\nn \\
+ \eta_8''\left(\frac{\p^2}{\p\eta_3'\p\eta_5'}
\,\hat{-}\frac{\p^2}{\p\eta_2'\p\eta_6'}\right)
+ \eta_8'''\left(\frac{\p^2}{\p\eta_2'\p\eta_6'''}
+\frac{\p^2}{\p\eta_3'\p\eta_5'}\right)
+ \eta_8''''\left(\frac{\p^2}{\p\eta_3'\p\eta_5'}
+\frac{\p^2}{\p\eta_3'''\p\eta_5'''}\right)
\ee
which acts on the space $\underline{{\cal C}_1}$ of dimension
${\rm dim}_q(\underline{{\cal C}_1}) = 4+4q^2=4qD$ as follows:

\bigskip

\centerline{
$
\underline{d_2}\ \downarrow\
\begin{array}{c||c|c|c|c|c|c|c|c|}
\underline{{\cal C}_1} & \theta_2'\eta_6'& \eta_2'\eta_6'&
 \theta_2'\eta_6'''& \eta_2'\eta_6'''& \eta_3'\theta_5'& \eta_3'\eta_5'&
\theta_3'''\eta_5'''&  \eta_3'''\eta_5'''\\
&&&&&&&&\\
\underline{d_2}\underline{{\cal C}_1} &\eta_2\theta_2'\eta_4&
\eta_2\eta_2'\theta_4 + \eta_8'\hat{-}\eta_8''& \eta_2\theta_2'\eta_4&
\eta_2\eta_2'\theta_4+\eta_8+\eta_8'''& 0& \eta_8''+\eta_8'''+\eta_8''''&
0&\hat{-}\eta_8+\eta_8'+\eta_8''''
\end{array}
$
}

\bigskip

\noindent
so that
\be
{\rm Ker}(\underline{d_2}) = \Big\{ \eta_3'\theta_5',\ \ \theta_3'''\eta_5''',\ \
\theta_2'(\eta_6'-\eta_6'''),\ \
 \eta_2'(\eta_6'-\eta_6''') + \eta_3'\eta_5' - \eta_3'''\eta_5''' \Big\}, \nn \\
{\rm dim}_q{\rm Ker}(\underline{d_2}) = 3+q^2,\ \ \ \ \ \ \ \ \
\underline{H_1}  = {\rm Ker}(\underline{d_2})/{\rm Im}(\underline{d_1})
=  \Big\{ \theta_3'\eta_5'\Big\}, \ \ \ \ \ \ \ \ {\rm dim}_q(\underline{H_1}) = 1
\ee
and
\be
{\rm Im}(\underline{d_2}) = \Big\{ \eta_2\theta_2'\theta_4,\ \
\eta_2\eta_2'\theta_4 + \eta_8 + \eta_8''', \ \
\eta_8 \hat{-} \eta_8' \hat{-} \eta_8'''',\ \
\eta_8''+\eta_8'''+\eta_8'''' \Big\},\nn \\
{\rm dim}_q{\rm Im}(\underline{d_2}) = q^{-1} + 3q, \ \ \ \ \
{\rm dim}_q{\rm Ker}(\underline{d_2}) + q\cdot{\rm dim}_q{\rm Im}(\underline{d_2})
= {\rm dim}_q({\cal C}_1) = 4+4q^2 = 4qD
\label{im4d2}
\ee
The combination $\eta_8 - \eta_8' \hat{+}\, \eta_8'' + \eta_8'''
\in {\rm Im}(\underline{d_2})$
as a linear combination of the two last entries in (\ref{im4d2}).

\subsection{Differential \underline{$d_3$}\ and the cohomology \underline{$H_2$}}

From
\be
\underline{K_3} =  p_6\frac{\p^2}{\p p_2'\p p_4}
\ooplus \underline{p_2}p_6\left(\frac{\p}{\p \underline{p_8}}
+ \frac{\p}{\p \underline{p_8'}}\right)
\ +\   p_6''\frac{\p^2}{\p p_2'\p p_4}
+ \underline{p_2}p_6''\left(\frac{\p}{\p \underline{p_8''}}
\ooplus \frac{\p}{\p \underline{p_8'''}}\right)
+ \nn \\
+  p_3''\underline{p_5''}\left(\frac{\p}{\p \underline{p_8'}}
+ \frac{\p}{\p \underline{p_8''}}
\ooplus \frac{\p}{\p \underline{p_8''''}}\right)
+  p_3\underline{p_5}\left(
\frac{\p}{\p \underline{p_8'''}} \ooplus \frac{\p}{\p \underline{p_8}}
\ooplus \frac{\p}{\p \underline{p_8''''}}\right)
\ee
we get:
\be
\underline{d_3} = -(\theta_6+\theta_6'')\left(\frac{\p^2}{\p\theta_2'\p\eta_4} +
\frac{\p^2}{\p\eta_2'\p\theta_4}\right)
- (\eta_6+\eta_6'')\frac{\p^2}{\p\eta_2'\p\eta_4} + \nn \\
\hat{-} \eta_2\theta_6\left(\frac{\p}{\p\eta_8} + \frac{\p}{\p\eta_8'}\right)
+ \eta_2\theta_6''\left(\frac{\p}{\p\eta_8''}\, \hat{-} \frac{\p}{\p\eta_8'''}\right)
+ \eta_3\theta_5\left(\frac{\p}{\p\eta_8'''} \, \hat{-}\frac{\p}{\p\eta_8}
\, \hat{-} \frac{\p}{\p\eta_8''''}\right)
+ \eta_3''\theta_5''\left(\frac{\p}{\p\eta_8''} + \frac{\p}{\p\eta_8'}
\, \hat{-} \frac{\p}{\p\eta_8''''}\right)
\ee
acts on the space $\underline{{\cal C}_2}$ of dimension
${\rm dim}_q(\underline{{\cal C}_2}) = q^{-1}+(2+5)q + q^3 = q(D^2+5)$:

\bigskip

\centerline{
$
\underline{d_3}\ \downarrow\
\begin{array}{c||c|c|c|c|}
\underline{{\cal C}_2}&\eta_2\theta_2'\theta_4&
\eta_2\theta_2'\eta_4& \eta_2\eta_2'\theta_4& \eta_2\eta_2'\eta_4 \\
&&&&\\
\underline{d_3}\underline{{\cal C}_2} &0& \eta_2(\theta_6+\theta_6'')&
\eta_2(\theta_6+\theta_6'')& \eta_2(\eta_6+\eta_6'')
\end{array}
$
}

\bigskip

\centerline{
$
\begin{array}{|c|c|c|c|c|}
\eta_8& \eta_8'& \eta_8''&\eta_8'''& \eta_8''''\\
&&&&\\
\hat{-}\eta_2\theta_6\hat{-}\eta_3\theta_5& \hat{-}\eta_2\theta_6+\eta_3''\theta_5''&
\eta_2\theta_6'' + \eta_3''\theta_5''& \hat{-}\eta_2\theta_6'' + \eta_3\theta_5&
\hat{-}\eta_3\theta_5\hat{-}\eta_3''\theta_5''
\end{array}
$
}

\bigskip

\noindent
Therefore
$$
{\rm Ker}(\underline{d_3}) = \Big\{ \eta_2\theta_2'\theta_4,\ \
\eta_2(\theta_2'\eta_4-\eta_2'\theta_4),\ \
\eta_2\eta_2'\theta_4+\eta_8+\eta_8''',\ \
\eta_8-\eta_8'-\eta_8'''',\ \ \eta_8''+\eta_8'''+\eta_8'''' \Big\},
$$
\vspace{-0.4cm}
\be
{\rm dim}_q{\rm Ker}(\underline{d_3}) =  q^{-1}+4q, \nn \\
\underline{H_2} = {\rm Ker}(\underline{d_3})/{\rm Im}(\underline{d_2})
=  \Big\{ \eta_2(\theta_2'\eta_4-\eta_2'\theta_4) \Big\}, \ \ \ \ \ \
{\rm dim}_q(\underline{H_2}) = q
\ee
and
\be
{\rm Im}(\underline{d_3}) = \Big\{\eta_2(\theta_6+\theta_6''),\ \
\eta_3\theta_5+\eta_3''\theta_5'',\ \ \eta_2\theta_6+\eta_3\theta_5,\ \
\eta_2(\eta_6+\eta_6'') \Big\},\nn \\
{\rm dim}_q{\rm Im}(\underline{d_3}) = 3+q^2, \ \ \ \ \
{\rm dim}_q{\rm Ker}(\underline{d_3}) + q\cdot{\rm dim}_q{\rm Im}(\underline{d_3})
= {\rm dim}_q({\cal C}_2) = q^{-1} + 7q + q^3
\ee

\subsection{Differential \underline{$d_4$}\ and the cohomologies \underline{$H_3$}\
and \underline{$H_4$}}

Finally,
\be
\underline{K_4} = p_3p_3''\left( \frac{\p}{\p p_6''}
\ooplus\frac{\p}{\p p_6}\right)
+\underline{p_2}p_3''\frac{\p}{\p \underline{p_5}}
\ooplus \underline{p_2}p_3\frac{\p}{\p \underline{p_5''}}
\ee
The corresponding differential
\be
\underline{d_4} =
-\theta_3\theta_3''\left(\frac{\p}{\p\theta_6''}\,\hat{-}\frac{\p}{\p\theta_6}\right)
- (\theta_3\eta_3'' \hat{-}\,\eta_3\theta_3'')
\left(\frac{\p}{\p\eta_6''}-\frac{\p}{\p\eta_6} \right)
+ \eta_2\theta_3''\frac{\p}{\p\eta_5} \hat{+}\, \eta_2\theta_3\frac{\p}{\p\eta_5''}
\ee
converts the $\underline{{\cal C}_3}$ of dimension
${\rm dim}_q(\underline{{\cal C}_3}) =  4+4q^2=4qD$:

\bigskip

\centerline{
$
\underline{d_4}\ \downarrow\
\begin{array}{c||c|c|c|c|c|c|c|c|}
\underline{{\cal C}_3} &\eta_2\theta_6& \eta_2\eta_6&
 \eta_2\theta_6''& \eta_2\eta_6''&
\theta_3\eta_5& \eta_3\eta_5& \theta_3''\eta_5''&\eta_3''\eta_5''\\
&&&&&&&&\\
\underline{d_4}\underline{{\cal C}_3}&
\hat{-}\,\eta_2\theta_3\theta_3''& \hat{-}\,\eta_2(\theta_3\eta_3''+\eta_3\theta_3'')&
\eta_2\theta_3\theta_3''& \eta_2(\theta_3\eta_3''+\eta_3\theta_3'')&
\eta_2\theta_3\theta_3''& \eta_2\eta_3\theta_3''&
\hat{-}\,\eta_2\theta_3\theta_3''& \hat{-}\,\eta_2\theta_3\eta_3''
\end{array}
$
}

\bigskip

\noindent
Thus
\be
{\rm Ker}(\underline{d_4}) = \Big\{ \eta_2(\theta_6+\theta_6''),\ \
\theta_3\eta_5 + \theta_3''\eta_5'',\ \
\eta_2\theta_6+\theta_3\eta_5,\ \
\eta_2\eta_6+\eta_3\eta_5-\eta_3''\eta_5'', \ \
\eta_2(\eta_6+\eta_6'') \Big\}, \nn \\
{\rm dim}_q{\rm Ker}(\underline{d_4}) = 3+2q^2, \nn \\
\underline{H_3} = {\rm Ker}(\underline{d_4})/{\rm Im}(\underline{d_3})
=  \Big\{ \eta_2\eta_6+\eta_3\eta_5-\eta_3''\eta_5'' \Big\},
\ \ \ \ \ \ {\rm dim}_q(\underline{H_3}) =  q^2
\ee

\noindent
The space
\be
{\cal C}_4 = \Big\{\eta_2\theta_3\theta_3'',\ \ \eta_2\theta_3\eta_3'',\ \
\eta_2\eta_3\theta_3'',\ \ \eta_2\eta_3\eta_3''\Big\}, \ \ \ \ \ \ \ \
{\rm dim}_q(\underline{{\cal C}_4}) =  q^{-1}+2q+q^3 = qD^2
\ee

\be
{\rm Im}(\underline{d_4}) = \Big\{ \eta_2\theta_3\theta_3'',\ \
\eta_2\theta_3\eta_3'',\ \
\eta_2\eta_3\theta_3'' \Big\}
\nn \\
{\rm dim}_q{\rm Im}(\underline{d_4}) = q^{-1}+2q, \nn \\
{\rm Coim}(\underline{d_4}) = \{\eta_2\eta_3\eta_3''\},
\ \ \ \ {\rm dim}_q(\underline{H_4}) =
{\rm dim}_q{\rm Coim}(\underline{d_4}) = q^3
\ee

\subsection{Reduced Jones superpolynomials for $4_1$}

Thus we have
\be
{\rm dim}_q(\underline{H_0})=q^{-1}, \nn\\
{\rm dim}_q(\underline{H_1})=1, \nn\\
{\rm dim}_q(\underline{H_2})=q, \nn\\
{\rm dim}_q(\underline{H_3})=q^{2}, \nn\\
{\rm dim}_q(\underline{H_4})=q^{3}
\ee
so that the reduced Jones superpolynomial is
\be
\underline{P}^{4_1}_\Box =
q^{-1} \frac{q^2}{(q^2T)^2} \sum_{i=0}^4 (qT)^i\cdot {\rm dim_q}(\underline{H_i})
= \frac{1}{q} \frac{q^2}{(q^2T)^2}\Big(q^{-1} + (qT)
+ q(qT)^2 + q^2(qT)^3 + q^3(qT)^4\Big) = \nn \\ =
\frac{1}{q^4T^2} + \frac{1}{q^2T} + 1 + q^2T + q^4T^2
\ee
what coincides with the value of the superpolynomial \cite{spfirst,IMMMf8}
\be
\boxed{
P^{4_1}_\Box(a|q|T) =
1+T^2a^2 + q^2T+\frac{1}{q^2T} + \frac{1}{T^2a^2}
}
\ee
at $a=q^2$.

\section{Conclusion}

This second part of review series explains the difference between
reduced and unreduced superpolynomials from the perspective of
Khovanov-Rozansky categorification approach.
We did not discuss the Chern-Simons-theory origin of this reduction,
as well as its (in)dependence on the choice of the "marked" edge
in the link diagram.
Instead, since for small knots reduced polynomials are considerably
simpler, we used this chance to describe cohomology calculus
in much more detail than it was done in \cite{DM}.

\section*{Acknowledgements}

Our work is partly supported by the
Ministry of Education and Science of the Russian Federation under the contract 8207,
by the grants
NSh-3349.2012.2,
RFBR-10-01-00538,
by the joint grants
90453-Ukr,  11-01-92612-Royal Society, 12-02-92108-Yaf-a
and by CNPq 400635/2012-7,
the Brazil National Counsel of Scientific and Technological Development.

\end{document}